\documentclass[onecolumn,authoryear]{els-mrw}

\usepackage{amsmath,amssymb,amsfonts,amsthm,makeidx,graphicx}
\usepackage{txfonts}
\usepackage{helvet}
\usepackage{wrapfig}
\usepackage{tabularx}

\newcommand*\aap{A\&A}

\newcommand*\aj{AJ}

\newcommand*\apj{ApJ}
\newcommand*\apjl{ApJL}

\newcommand*\apjs{ApJS}

\newcommand*\mnras{MNRAS}

\newcommand*\nat{Nature}

\newcommand*\pasp{PASP}

\newcommand*\ssr{Space~Sci.~Rev.}

\def\ms{m~s$^{-1}$}
\def\js{J~s$^{-1}$}
\def\msunyr{M$_{\odot}$\,yr$^{-1}$}
\def\msun{M$_{\odot}$}
\def\rsun{R$_{\odot}$}
\def\lsun{L$_{\odot}$}

\def\tiff{\iso{44}Ti}
\def\crfe{\iso{48}Cr}
\def\nifsev{\iso{57}Ni}
\def\nifs{\iso{56}Ni}
\def\cofs{\iso{56}Co}

\newcommand{\iso}[2]{\ensuremath{^{#1}\rm{#2}}}

\def\one{{\,\sc i}}
\def\two{{\,\sc ii}}
\def\three{{\,\sc iii}}

\def\oidoub{[O\one]\,$\lambda\lambda$\,$6300,\,6364$}

\begin{document}

\chapter{Interacting Supernovae}\label{intsn}

\author[1]{Luc Dessart}

\address[1]{\orgname{Institut d'Astrophysique de Paris}, \orgdiv{CNRS-Sorbonne Universit\'e}, \orgaddress{98 bis boulevard Arago, F-75014 Paris, France.}}

\articletag{Chapter Article tagline: update of previous edition,, reprint..}

\maketitle

\begin{glossary}[Glossary and nomenclature]
\begin{tabular}{@{}lp{34pc}@{}}
Stellar explosion & Process by which the hydrostatic equilibrium of a star is destroyed through the sudden release of an energy that far exceeds the star's gravitational binding energy. This energy release is thought to arise from the gravitational collapse of the degenerate core in massive stars or from a thermonuclear runaway in a white dwarf star.  \\ 
Supernova (SN) & Generic name referring to a wide diversity of stellar explosions and typically characterized by a week- to month-long  high luminosity phase of order 10$^8$--10$^9$ times the luminosity of the sun.\\
SN light curve & Photometric or brightness evolution of SNe in a specific region of the electromagnetic spectrum. Different filter passpands cover from the ultraviolet to the infrared. \\ 
SN spectra      &  Quantity describing the precise distribution (rather than the broad variations captured by photometry) of the radiative flux with wavelength. Spectra convey information on ejecta velocity, elemental composition, temperature, ionization, densities, and allow inferences on a variety of physical processes. In SN observations, the typical spectral resolution $\lambda/\Delta\lambda$ used is $\sim$\,1000, corresponding in velocity space to a resolution of $\sim$\,3$\times$10$^5$\,\ms. \\ 
 Type II SN    & Supernova exhibiting spectral lines of hydrogen.\\
Type IIP SN  & Type II SNe exhibiting a roughly constant optical brightness (`P' stands for plateau) for about 100\,d after explosion. \\
Type IIn SN  & Type II SNe characterized by spectral lines that are significantly narrower than normally observed in SNe.\\
Type Ib SN   & Supernova exhibiting spectral lines of helium but not of hydrogen.\\
Type Ibn SN & Counterparts of SNe IIn but for Type Ib SNe exhibiting relatively narrow He lines. \\
Type Ic SN   & Supernova exhibiting no spectral lines of hydrogen, helium, nor of silicon.  \\
Type Icn SN & Counterparts of Type Ibn SNe that exhibit relatively narrow C and O lines rather than He lines.\\
Electron scattering & This process describes the deflection or scattering of radiation by free electrons in an ionized gas. This scattering, which is generally assumed to be isotropic, comes with a shift at frequency $\nu$ on the order of $\nu V_{\rm th,e}/c$, where $V_{\rm th,e}$ is the velocity of these (thermal) free electrons and $c$ is the speed of light. At an electron temperature of 10\,000\,K,  $V_{\rm th,e}$ is of the order of 5.5$\times$\,10$^5$\,\ms. The electron-scattering opacity $\sigma_{\rm e}$ is independent of wavelength and is equal to 6.65$\times$\,10$^{-29}$\,m$^2$.   \\
Optical depth & A measure of the absorption or scattering affecting radiation travelling along a given path length. For example, the radial electron scattering optical depth integrated inwards from infinity down to a radius $R$ is defined as $\tau_{\rm e} = \int_R^\infty N_{\rm e} \sigma_{\rm e} dr$, where $N_{\rm e}$ is the density of free electrons at locations along that radial direction. \\
Photosphere & Location in an ejecta where the radial optical depth integrated from infinity is equal to 2/3. At the electron-scattering photosphere corresponds a radius $R_{\rm ph}$ defined through  $\int_{R_{\rm ph}}^\infty N_{\rm e} \sigma_{\rm e} dr=$\,2/3. The photosphere separates the regions that are optical thin and optically thick.\\
Shock breakout &  Phenomenon describing the emergence of the SN shock at the surface of the exploding, progenitor star. \\
Radiative diffusion & Process describing the transport of radiation in an optically thick medium.\\
Radiative precursor & Short-lived phenomenon corresponding to the release of trapped radiation as the SN shock approaches the stellar surface. A radiative precursor forms when the shock crossing time $\Delta R/V_{\rm sh}$ through the remaining layers of the star of thickness $\Delta R$ equals the radiative-diffusion time through those same layers and given by $\tau \Delta R / c$. This precursor forms when $\tau = c/V_{\rm sh}$, which is about 10--30 for core-collapse supernovae.\\
Photospheric phase & Phase during which the SN ejecta are optically thick to electron scattering. \\
Nebular phase & Corresponds to the subsequent phase when the ejecta are transparent. \\
Spectral line broadening & The intrinsic width of spectral lines is essentially set by the thermal velocity of the associated atoms and ions in the spectrum formation region, and thus of a few 10$^3$\,\ms.  Instead, spectral lines recorded by a distant observer exhibit a significant broadening. It can arise from the large velocities of the emitting regions relative to the observer. It can also arise through scattering with free electrons, both because of their relatively large thermal velocities (this redistribution in frequency or wavelength is random and thus tends to produce a symmetric line profile when acting alone) and because of the large ejecta velocities (scattering with free electrons elsewhere in the ejecta causes a systematic redshift of line photons as recorded by a distant observer at rest). SN spectral line profiles are typically broadened by a few 10$^{6}$\,\ms, and are affected by both thermal motions and bulk motions, although one generally dominates over the other.
\end{tabular}
\end{glossary}

\begin{abstract}[Abstract]
Modern photometric surveys of the sky suggest that many, perhaps most SNe associated with the explosion of massive stars are influenced at an appreciable level by their interaction with circumstellar material (CSM). The photometric and spectroscopic diversity of these transients point to a wide range of CSM properties in terms of mass, extent, composition, and location relative to the exploding star, suggesting progenitors that cover from standard to the most extreme mass loss rates. Surveys at high-cadence catch massive stars at shock breakout and inform us on the immediate mass loss history before core collapse. In contrast, long-term monitoring of these SNe cover the transition to the birth of a SN remnant and document the progenitor mass loss that took place centuries to millennia before explosion. Interacting SNe are therefore not just extraordinary astrophysical laboratories to study radiation-dominated shocks and probe the distant Universe, they also open the path to novel and fundamental studies on stellar evolution, stellar stability, or mass loss in single and binary massive stars. 
\end{abstract}

\begin{BoxTypeA}[]{Key points}
\begin{itemize}
\item Interaction between ejecta and CSM seems to be universal amongst massive star explosions. It dominates the dynamical evolution and shapes the radiative characteristics in strongly interacting events (e.g., Type IIn SNe), but it likely affects the properties of all SNe at some reduced level or at specific epochs. 
\item Interaction between ejecta and CSM is a powerful engine for the transformation of ejecta kinetic energy into radiative energy. Because kinetic energy is so abundant in SN ejecta, this process can produce super-luminous transients. Inherently tied to a shock, the interaction process can also lead to emission across the electromagnetic spectrum, from X-rays to ultraviolet, optical, infrared, and even radio wavelengths.
\item Most interacting SNe today are discovered with ground-based telescopes in the optical range. When the CSM is hydrogen rich, their early spectra exhibit narrow, symmetric emission lines broadened in the process of scattering with free electrons. This is in stark contrast with noninteracting SNe and their wide, asymmetric lines broadened as a result of the large ejecta velocities. Interacting SNe offer an extraordinary diversity in spectral properties, in particular in terms of line profile morphology and evolution, or levels of ionization, testifying for their exotic nature and origin.
\item Interacting SNe provide complementary means to study the evolution of massive stars, their mass loss properties in the final stages of evolution prior to core collapse and explosion. Because of their large luminosity, they can be used to probe the distant universe. Interaction with CSM offers as well a means to monitor the transition of SNe into young SN remnants.
\end{itemize}
\end{BoxTypeA}

\section{Introduction}\label{intsn_intro}

All massive stars below about 140\,\msun\ end their lives with the formation of an iron core that collapses to a protoneutron star (e.g., \citealt{whw02}, \citealt{woosley_ppsn_17}). The fate of the overlying layers is determined by the interplay between accretion onto the compact remnant and neutrino energy deposition in the infalling material, both orchestrated by numerous fluid instabilities (e.g., \citealt{burrows_rev_21}). When successful, this mechanism leads to a core-collapse supernova (SN) with a peak luminosity of a billion times that of the Sun and an explosion energy comparable to several times the binding energy of the matter outside the collapsed core (e.g., \citealt{sukhbold_ccsn_16}).

Until recently, the diversity of SNe was generally attributed to the nature of the progenitor - its mass, its extent, its rotation rate, the compactness of the stellar core at the time of explosion, or its belonging to a binary system. While critical in determining the outcome of the explosion, these important features only partly explain the emerging panoply of SN observations. Indeed, the considerable diversity revealed by high-cadence photometric surveys of the sky (e.g., ZTF, \citealt{ZTF}; ASAS-SN, \citealt{ASASSN}; ATLAS, \citealt{atlas_18}) and by spectroscopic monitoring of these transients suggest that the recent, pre-explosion history of the star and its environment are a major component in shaping SN light curves and spectra: massive stars typically explode not in a vacuum but instead in circumstellar material (CSM) produced by that very same star (and possibly its companion for interacting binaries) during its evolution prior to explosion. The current census indicates that about 10\% of core-collapse SNe are classified as interacting (e.g., either Type IIn or Ibn; \citealt{perley_rates_20}). SN observations thus provide unprecedented constraints on the properties of SN progenitors and their evolution. 

This chapter is structured as follows. Section~\ref{intsn_obs_diversity} presents the observational evidence for interaction in SNe from exploding massive stars.  Section~\ref{intsn_proc} presents the various processes that may be at the origin of the CSM involved in such ejecta/CSM interactions. The light curve diversity of interacting SNe is explored more quantitatively with the help of radiation-hydrodynamics simulations in Section~\ref{intsn_rhd} -- such numerical simulations are used throughout the chapter for comparison to data. Section~\ref{intsn_ref_case} explores in detail the spectral evolution of a representative case and what each phase of evolution corresponds to in terms of dynamics and radiative transfer. This knowledge is then used to address the diversity of interacting SNe in terms of confined CSM (short-lived interactions), extended CSM (long-lived interactions), or variations in CSM mass relative to ejecta mass (quenched interactions; Section~\ref{intsn_iin_diversity}). Section~\ref{intsn_other} turns to more exotic and rarer interactions in which the ejecta or the CSM are free of hydrogen, and sometimes of helium, while Section~\ref{intsn_dust} discusses the evidence for dust formation in interacting SNe. Section~\ref{intsn_conc} wraps up with some concluding remarks and an outlook for the topic of interacting SNe. Further information on interacting SNe is provided in a number of recent reviews (see, e.g., \citealt{blinnikov_rev}; \citealt{chevalier_fransson_rev_17}; \citealt{smith_rev}).

\section{The observational landscape of interacting supernovae}
\label{intsn_obs_diversity}

\begin{figure}[t]
\centering
\includegraphics[width=0.49\textwidth]{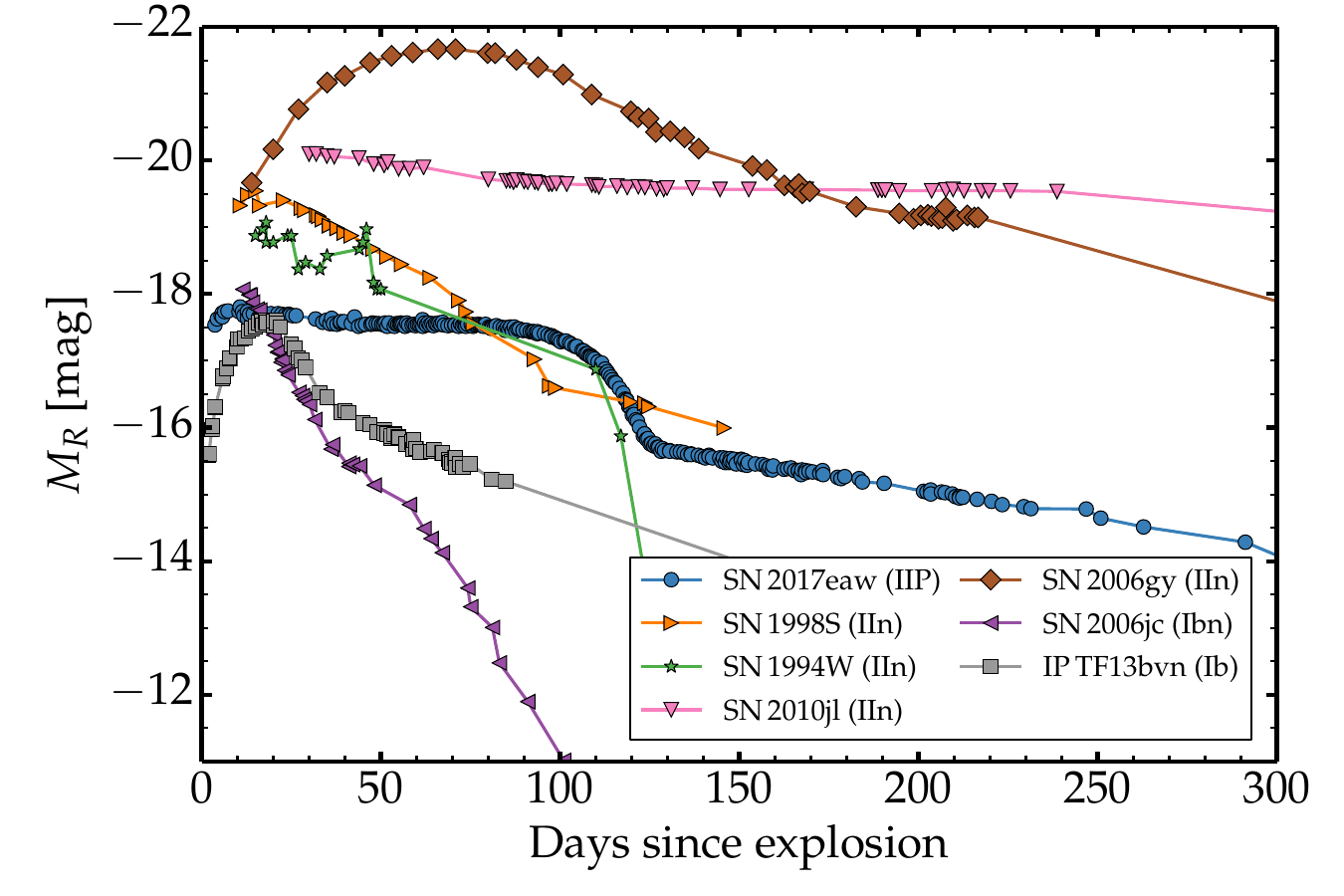} 
\includegraphics[width=0.49\textwidth]{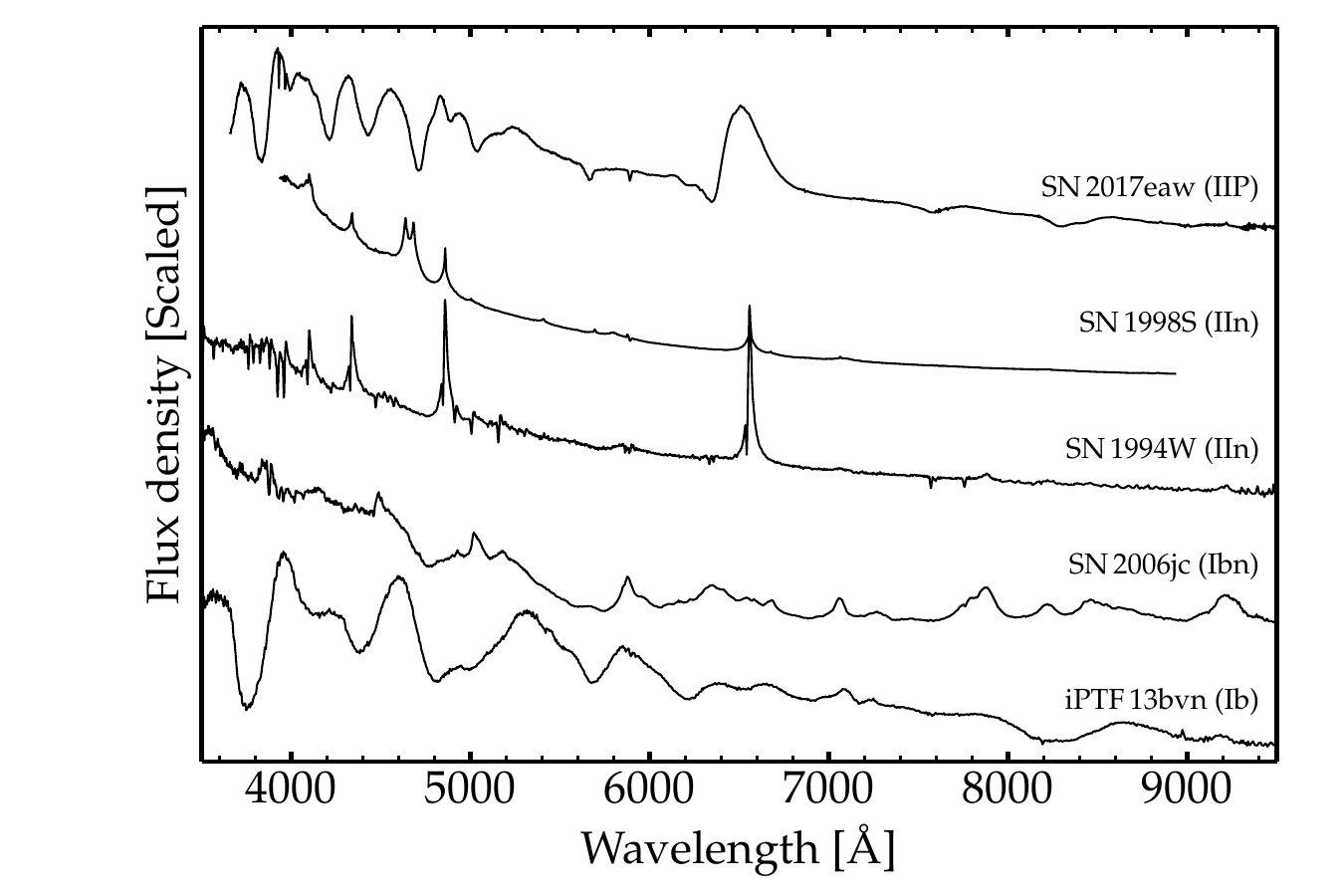} 
    \caption{Diversity in the evolution of the optical brightness and maximum-light spectra for a sample of core collapse SNe with and without interaction. Left: We show the $R$-band magnitude for Type IIn SNe 1994W, 1998S, 2006gy, and 2010jl, for Type II-P SN 2017eaw, for Type Ibn SN\,2006jc and for Type Ib SN\,iPTF13bvn. Right: Sample of optical spectra taken around optical-brightness maximum and shown for the SN types shown at left. Data sources are given in the text.}
\label{intsn_fig_obs_div}
\end{figure}

Photometry and spectroscopy are the two main types of information on SNe. Photometry and multi-band light curves reveal the rate, the amount and the broad wavelength distribution of the radiation that escapes through the photosphere of optically-thick ejecta. In explosions of red-supergiant stars, the optical (or $R$-band) light curve is typically characterized by a near constant brightness of about $-17$\,mag, which leads to a classification as Type II-Plateau (or IIP) SN. In explosions of more compact, stripped-envelope progenitors, the significant cooling from expansion produces a brightness that is initially low but rises to a peak of comparable magnitude on a 20-d timescale because of radioactive decay heating -- the light-curve characteristics (i.e., rise time and duration) are controlled by radiative diffusion and ejecta expansion. The main unstable isotope is \nifs, which is explosively produced with a typical mass of order 0.01\,\msun. A representative ejecta kinetic energy $E_{\rm kin}$ for these standard core-collapse SNe is about 10$^{44}$\,J (see, e.g., \citealt{pejcha_prieto_sn2p_15}), whereas the total (aka bolometric)  time-integrated luminosity $E_{\rm rad}$ is only about one per cent of that. Two representative examples of Type IIP and Ib SNe are given in the left panel of Fig.~\ref{intsn_fig_obs_div} for SN\,2017eaw \citep{buta_keel_17eaw_19,vandyk_17eaw_19,weil_17eaw_20} and SN\,iPTF13bvn \citep{fremling_sesn_14}.

In contrast, when the SN expands within some CSM, the ejecta are decelerated by this external mass buffer, leading to an extraction of ejecta kinetic energy and its conversion into radiative energy. Because  $E_{\rm kin}/E_{\rm rad}\sim$\,100 in standard (i.e., noninteracting) SNe, the extraction of 1\% of $E_{\rm kin}$ can double $E_{\rm rad}$, the extraction of 10\% of $E_{\rm kin}$ can increase $E_{\rm rad}$ by a factor of ten, and the total exhaustion (i.e., complete deceleration of the ejecta) can increase $E_{\rm rad}$ by a factor of a hundred. The interacting Type II SNe (aka Type IIn SNe) suggest that this is precisely what happens in nature, yielding SNe that span from a modest (SN\,1994W, \citealt{sollerman_etal_94w}, \citealt{chugai_94w_04}; SN\,1998S \citealt{fassia_98S_00}, \citealt{leonard_98S_00}) to a huge enhancement (SN\,2010jl, \citealt{fransson_10jl}; SN\,2006gy, \citealt{smith_06gy_08}) in both peak luminosities (or brightness) and time-integrated bolometric luminosity $E_{\rm rad}$ (left panel of Fig.~\ref{intsn_fig_obs_div}). In the Type Ibn SN\,2006jc \citep{foley_06jc_07}, the peak luminosity is also large although not so well reflected by the $R$-band magnitude because a significant fraction of the flux falls at shorter wavelength, typically below 5500\,\AA\ and into the ultraviolet. The fast declining light curve is, however, genuine and arises from the much lower ejecta mass and optical depth, leading to the release of stored radiative energy on a short timescale. 

A second, but critical aspect is that this CSM influences the radiation from the exploding star. In the absence of CSM, the radiation from the fast-moving outer ejecta layers, where the photosphere normally resides at early times, gives rise to broad, Doppler-broadened line profiles at early times (e.g., as observed in SNe 2017eaw and iPF13bvn; see right panel of Fig.~\ref{intsn_fig_obs_div}). However, in the presence of CSM, this radiation is reprocessed by the unshocked, slower moving CSM, and gives rise to narrow line profiles (the reason for the ``n'' suffix in Type IIn, Ibn, or Icn SNe), as first evidenced with SN\,1983K (\citealt{niemela_etal_85}; see also \citealt{schlegel_iin_90}).  Depending on the CSM composition, different spectral signatures are observed. With an hydrogen-rich CSM, H\one\ spectral lines like H$\alpha$ exhibit a narrow core and extended symmetric `wings' (e.g. SNe\,1994W or 1998S).  With an H-deficient but He-rich CSM, spectral lines of He\one\ at 5876 or 7065\AA\ are produced (e.g., SN\,2006jc). Such peculiar signatures are also seen in some Type Ic, as well as Type Ia SNe, and these events will be discussed in Section~\ref{intsn_other}.

This external material may originate from a variety of processes, which are discussed in the next section. Observationally, there is evidence for phases of enhanced brightness or variability prior to the discovery of interacting SNe, suggesting that a fraction of Type IIn and Ibn SNe experience individual or repeated outbursts in the months to years before explosion (see, e.g.,  \citealt{strotjohann_presn_21}; \citealt{jacobson_galan_20tlf_22}; \citealt{pasto_06jc_07}). Surprisingly, there is also evidence of the opposite with objects that clearly show interaction with CSM but appear quiescent in the years preceding explosion (see., e.g., SN\,2023ixf;  \citealt{ransome_23ixf_24}).

Constraints from progenitor detection (e.g., \citealt{galyam_leonard_05gj_09}) remain equivocal for interacting SNe since they cannot be reliably connected to a pre-SN mass or an initial mass on the main sequence. Indeed, the progenitor luminosity may already depart from the steady power released by a star in hydrostatic equilibrium (as modeled in stellar evolution codes). It may instead reflect the anomalous, erratic, unstable nature of a star already in a phase of intense mass loss and erupting.  SN\,2020tlf, whose estimated progenitor mass is around 12\,\msun, had the luminosity of a 60\,\msun, `quiescent' star just before explosion \citep{jacobson_galan_20tlf_22}.

\section{Physical processes at the origin of circumstellar material around exploding massive stars}
\label{intsn_proc}

  A variety of processes may be at the origin of the CSM present in interacting SNe. We first discuss the processes that apply to all stars and then turn to those that arise exclusively in stars having a companion (aka binary stars).

   \subsection{Mass loss mechanisms related to all massive stars}
       \subsubsection{Stellar winds}  
\label{intsn_wind}

The transfer of radiation momentum to matter is at the origin of the stellar winds identified around massive stars \citep{cak}. The rate of mass loss is typically on the order of 10$^{-6}$ to 10$^{-5}$\,\msunyr, and the asymptotic velocity of this wind scales with the surface gravity: compact, H-deficient stars (e.g., Wolf-Rayet stars) have fast winds of order 10$^6$\,\ms, whereas extended, generally H-rich stars (e.g., red supergiant stars) have slow winds of order 10$^4$\,\ms. The acceleration from zero to the wind asymptotic velocity  occurs over an extended region beyond the hydrostatic surface. This region may thus be relatively dense and act as a buffer for the incoming SN shock. In cool, supergiant stars, this buffer may become ``overloaded" over time if the star pulsates or if the wind driving is inefficient (which would cause fallback of the lifted material). Thus, by the end of the red-supergiant phase, which typically lasts a million years, the star  may be surrounded by a cocoon of material.

       \subsubsection{Wave excitation by core convection}
\label{intsn_wave}

Through the  final burning stages (i.e., C, Ne, O, and Si burning), the nuclear power released in the core of massive stars increases dramatically from 10$^7$ up to 10$^{11}$\,\lsun. Most of this power is directly radiated away in the form of neutrinos, while the rest generates convection, which excites waves that travel into the envelope of the star where they are absorbed. Numerical simulations suggest that this wave excitation power may reach of order 10$^{7}$\,\lsun, which is typically a hundred time greater than the star's luminosity, causing the expansion of the envelope and possibly boosting the wind mass loss \citep{quataert_shiode_12,fuller_rsg_17}. In a binary system, this process could have further implications (see Section~\ref{intsn_bin}).

 \begin{wraptable}{r}{0.5\textwidth} %% Coding for Non-floating table
 \vspace{-1.4cm}
\caption{Properties of the ejecta and CSM used as initial conditions for the radiation hydrodynamics calculations of \citet{D15_2n} whose bolometric light curves are shown in Fig.~\ref{intsn_fig_lbol}. In all cases, the mass of the SN ejecta are 9.8\,\msun; the CSM moves at 10$^5$\,\ms\ and extends out to 10$^{14}$\,m. The rise time of the light curve is the time difference between 1\% of and the bolometric maximum. Numbers in parenthesis represent powers of ten.\label{tab_rhd}}
\begin{tabular}{l@{\hspace{1mm}}c@{\hspace{2mm}}c@{\hspace{2mm}}
     c@{\hspace{2mm}}c@{\hspace{2mm}}c@{\hspace{2mm}}c@{\hspace{2mm}}}
 \toprule
%\begin{tabular}{lcccccc}
%\hline
model        &$E_{\rm kin, SN}$ &  $M_{\rm CSM}$ & $\dot{\rm M}_{\rm CSM}$ & $L_{\rm bol, peak}$  &  $t_{\rm peak}$  & $\int L dt$  \\
             & [10$^{44}$\,J]       & [\msun] & [\msunyr]  &    [J\,s$^{-1}$]  & [d]    & [10$^{44}$\,J] \\
% \colrule
\hline
X            &    1   &      2.89   &   0.1    & 3.024(36)  &    19.4 &  0.32    \\
Xe3          &    3   &      2.89   &   0.1    & 1.204(37)  &    15.7 &  0.88    \\
Xe3m6        &    3   &     17.31   &   0.6    & 2.080(37)  &    55.7 &  2.05    \\
Xe3m6r       &    3   &     26.73   &   0.6    & 1.818(37)  &    68.3 &  2.13    \\
Xe10         &    10  &      2.89   &   0.1    & 6.399(37)  &    12.7 &  2.92    \\
Xe10m6       &    10  &     17.31   &   0.6    & 1.091(38)  &    34.2 &  6.89    \\
Xm3          &    1   &      8.66   &   0.3    & 3.906(36)  &    47.9 &  0.49    \\
Xm6          &    1   &     17.31   &   0.6    & 4.751(36)  &    77.5 &  0.63    \\
\hline
% \botrule
\end{tabular}
\end{wraptable}

       \subsubsection{Nuclear flashes in 9 and 11\,\msun\ stars}
\label{intsn_flash}

In low mass massive stars (i.e., mass range between 9 and 11\,\msun), the ultimate stages of nuclear burning in the core take place at very high density and possibly under degenerate conditions. Rather than a slow and steady burning, combustion may be violent and lead to the release of energy on a very short time scale. This energy cannot be all transported by neutrinos, nor by photons or even convection, so the pressure ramps up, a shock forms, which subsequently crosses the envelope of the star. In supergiant stars, this can cause the complete ejection of the stellar envelope. Being intimately associated with the final stages of 9--11\,\msun\ stars, this phenomenon offers a natural synchronization between envelope ejection and the final explosion of the star. This process, relevant for the lighter massive stars, has been discussed in detail by \citet{WH15}.

\begin{wrapfigure}{R}{0.5\textwidth}
  \vspace{-0.8cm}
\begin{center}
\includegraphics[width=0.49\textwidth]{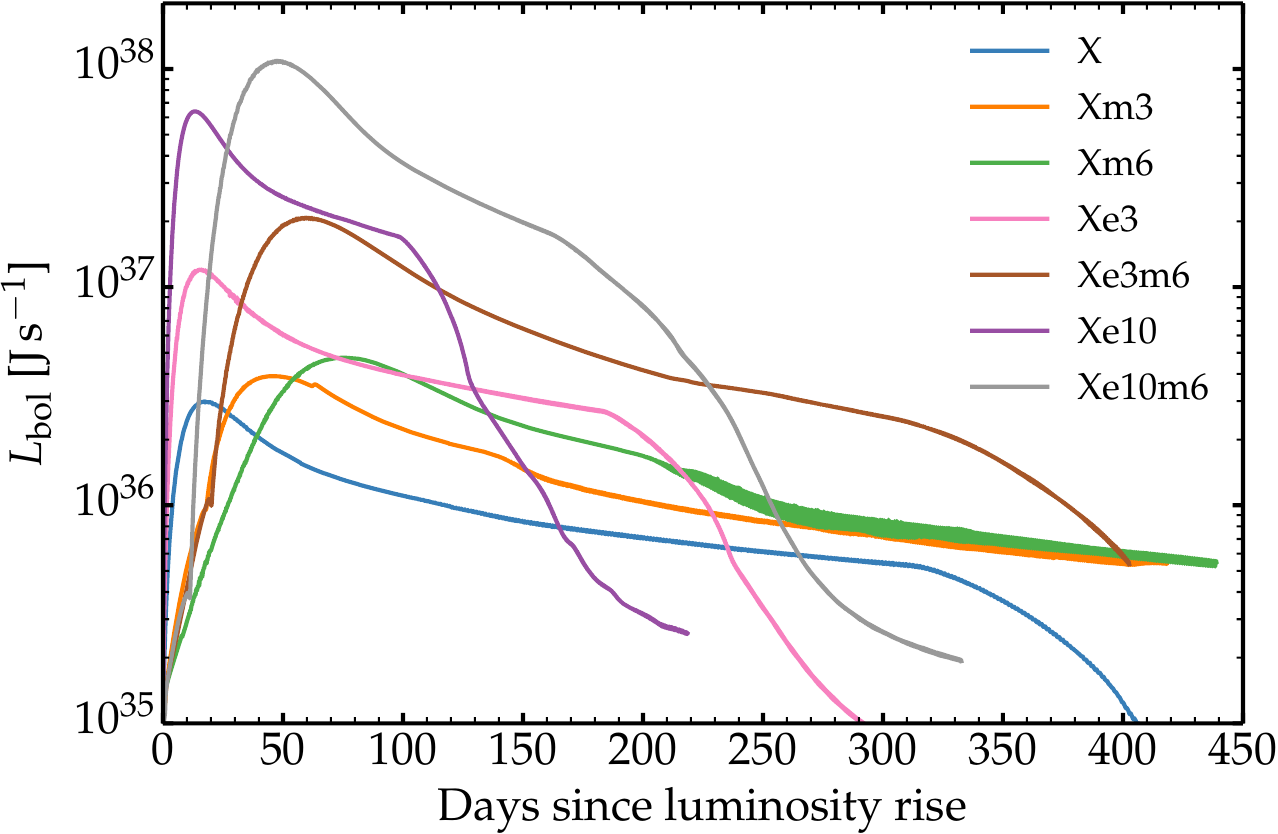} 
\end{center}
\vspace{-0.5cm}
    \caption{Illustration of the light curve diversity obtained with radiation-hydrodynamics simulations of a range of ejecta and CSM configurations (see Table.~\ref{tab_rhd}). Here, variations in the ejecta kinetic energy and the CSM mass lead to strong modulations of the rise time to peak, the luminosity at peak, or the duration of the high brightness phase.}
   \label{intsn_fig_lbol}
\end{wrapfigure}

      \subsubsection{Pulsational pair instability in high-mass stars}
\label{intsn_ppisn}

At the upper end of the mass range for massive stars, a unique instability takes place. The huge temperatures in the stellar interiors are conducive to the production of electron-positron pairs from photon annihilation. Because such cores are supported by radiation pressure, the formation of those pairs leads to a pressure deficit, inducing the collapse of the core, a compressional heating, and a thermonuclear runaway. For stars that form an He core between $\sim$\,30 and $\sim$\,60\,\msun, this instability is quenched by the cooling from expansion and avoids total disruption. Instead, the star initiates a cycle of pulses separated by quiescent phases of relaxation back to hydrostatic equilibrium.  Because the first pulses tend to eject more mass and be less energetic, material from later pulses eventually collide with material from previous pulses. To produce a luminous optical transient, the collision between the two pulses must take place within about 10$^{14}$\,m -- a larger distance would  likely imply a more optically-thin configuration and emission of shock power in the ultraviolet or X-ray ranges. It also restricts suitable configurations to a range of pulse delays and expansion rates. For example, long interpulse delays would imply interactions at very large distances with modest powers. At the other end, short delays would make the event appear as a single burst rather than an interaction. 

   \subsection{Mass loss mechanisms associated exclusively with binary massive stars}
\label{intsn_bin}

In addition to the processes described above, stars that reside in a binary system with an initial orbital separation smaller than about 1000\,\rsun\ will experience one or multiples phases of intense mass loss when they fill their Roche lobe.  This surface corresponds to the equipotential that crosses the first Lagrange point of null gravity in between the two stars. During such a Roche lobe overflow phase, mass is transferred to the companion star, lost by the binary system, or both. A plethora of scenarios are possible with such interacting binaries since the intrinsic complications of single-star evolution are extended by the possibilities of having different primary and secondary masses as well as different initial orbital separations. In the context of interacting SNe, stars that undergo Roche-lobe overflow shortly (i.e. few 10,000\,yr) before core collapse, thus during or after the He burning phase, are good candidates for the production of CSM that resides close to the exploding star \citep{podsiadlowski_92,ercolino_bin_23}.

Another possible scenario leading to interacting SNe is common envelope evolution in a binary system. This may occur in a variety of ways, for example if both stars fill there Roche lobe at the same time, if the mass transfer rate onto the companion is excessively large, or if a compact object is `swallowed' by the envelope of its companion when it becomes a red-supergiant star \citep{chevalier_ce_12}. The properties of the resulting interacting SN will depend strongly on the location of the CSM when the faster-evolving binary component eventually explodes.

\section{Modeling of the light curve diversity of interacting supernovae}
\label{intsn_rhd}

Figure~\ref{intsn_fig_lbol} illustrates the bolometric light curves resulting from the simulation of a variety of ejecta/CSM interaction configurations, involving different kinetic energies for the ejecta and masses (or mass loss rates) for the CSM (Table~\ref{tab_rhd}) -- these results are produced by radiation-hydrodynamics simulations that solve for the gas and radiation properties as a function of time \citep{D15_2n}. The origin of this emerging radiation is the power released at the shock $L_{\rm sh}$ given approximately as $L_{\rm sh} = 2 \pi R^2 \rho_{\rm CSM} V_{\rm sh}^3$, where $\rho_{\rm CSM}$ is the CSM density ahead of the shock and $V_{\rm sh}$ is the shock velocity (see, e.g., \citealt{moriya_lbol_13}). If the CSM arises from a steady-state wind with velocity $V_\infty$ and mass loss rate $\dot{M}$, this expression becomes $L_{\rm sh} = (\dot{M}/2V_\infty) V_{\rm sh}^3$. A large fraction of this shock power is thermalized by the CSM and transformed into lower energy photons that diffuse through the CSM, leading to a delayed peak in luminosity (this delay is greater for more massive and extended CSM). With time passing, the CSM optical-depth above the shock decreases, and the luminosity progressively converges to the value of the shock power. In all simulations, the sharp drop at late times occurs when the shock leaves the dense part of the CSM. As long as the shock is strongly embedded within the CSM, the bulk of the escaping radiation emerges in the UV and optical ranges. If or when this CSM is tenuous, the thermalization of this shock power is inefficient and high-energy radiation may escape, in particular in the form of X-rays (see, e.g., the recent review by  \citealt{chevalier_fransson_rev_17}), but also as radio emission (radio light curves are one additional source of information on interacting supernovae, not discussed here -- for information, see, e.g., \citealt{chandra_radio_rev}). In the interaction `engine', whatever is lost in kinetic energy comes out as radiation (i.e., the loss in $E_{\rm kin, SN}$ equals $\int L dt$, also named $E_{\rm rad}$ earlier on). 

\section{The various phases in the evolution of an interacting supernova}
\label{intsn_ref_case}

Below, we summarize the salient features characterizing the main phases of evolution of the interaction between ejecta and CSM. Each phase may vary in duration reflecting the differences in CSM properties (e.g., location, mass, extent, composition), ejecta properties (e.g., total mass, explosion energy, \nifs\ mass). Many possible combinations of CSM and ejecta properties lie behind the observed diversity of light curves and spectra of interacting SNe. 

\subsection{Shock breakout and the CSM phase}\label{intsn_sbo_csm}

  The explosion of a massive star corresponds to a sudden (i.e., 1s-long) energy release comparable to several times the binding energy of the envelope overlying the newly-born neutron star. A powerful shock quickly forms and crosses the envelope of the star, in about a minute in a Wolf-Rayet star and in about a day in a red-supergiant star. As the shock approaches within an optical depth of 10--30 of the surface, the radiation stored behind the shock starts `leaking', and diffuses through the remaining layers of the star. Put differently, the radiation ramdom-walks through the remaining optically-thick layers $\Delta R$ of the star  on a timescale $\tau \Delta R / c$ that is shorter than the time $\Delta R / V_{\rm sh}$ that it takes the shock to cross those same layers. This `shock breakout' is the first electromagnetic signature of the exploding star. One also talks of a radiative precursor because the radiation precedes the incoming shock. This process is analogous to seeing lightning before hearing thunder, which arises as well from the contrast between the speed of light and the speed of sound in the earth atmosphere. 
  
\begin{wrapfigure}{R}{0.5\textwidth}
\vspace{-0.9cm}
\begin{center}
\includegraphics[width=0.48\textwidth]{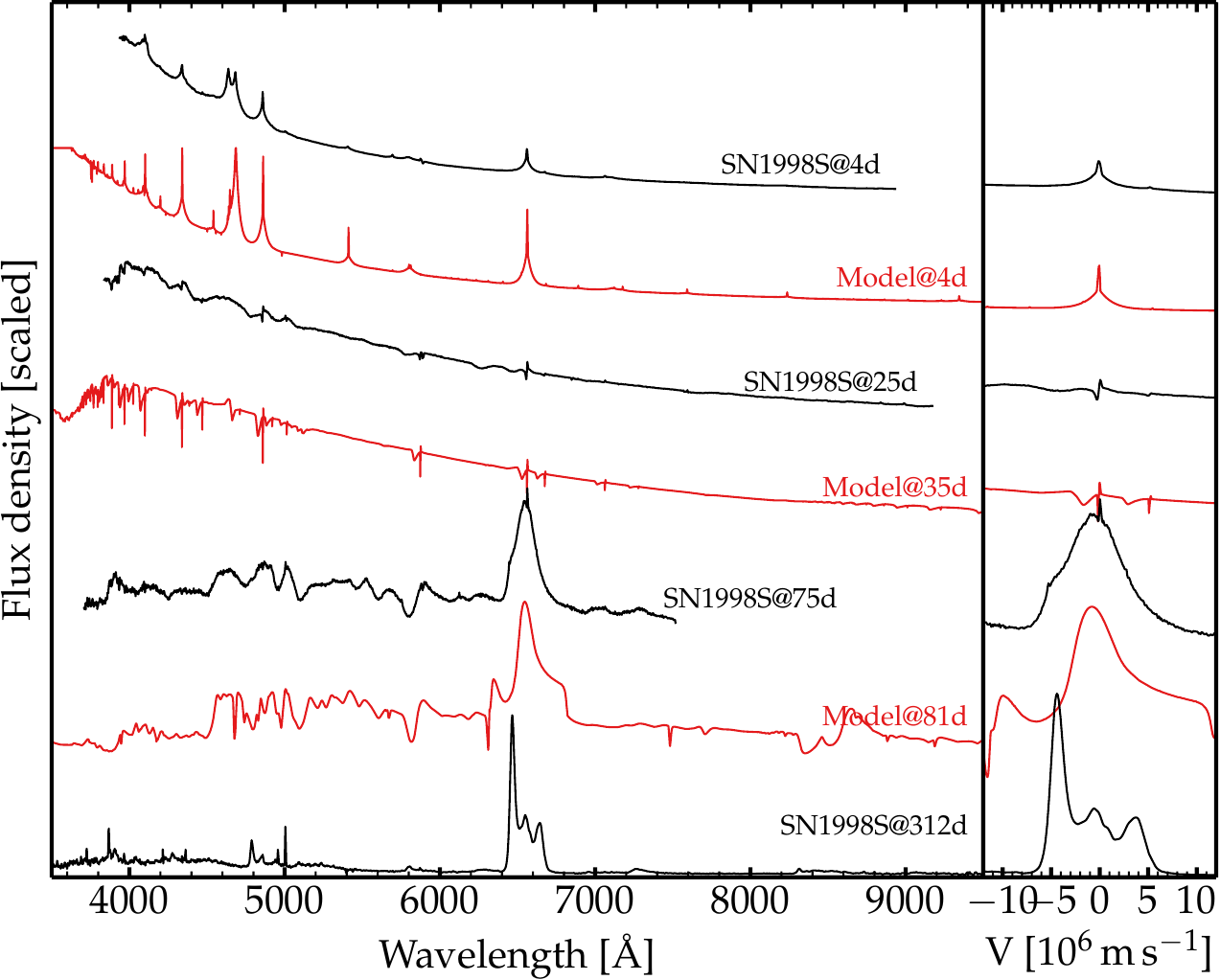} 
\end{center}
\vspace{-0.4cm}
    \caption{Illustration of the spectral evolution (left panel) of SN\,1998S (red) and a model counterpart as the ejecta/CSM interaction evolves through the `CSM phase' (example at 4\,d), the `CDS phase' (at 25-35\,d), the ejecta phase (at 75-81\,d)and the nebular phase (at 312\,d). Dramatic changes appear in the morphology of lines, in particular of H$\alpha$, which is shown in velocity space for the same epochs in the right panel. Model spectra are taken from \citet{D16_2n}, \citet{dessart_csm_22}, and \citet{dessart_wynn_23}. See Section~\ref{intsn_ref_case} for discussion.}
   \label{intsn_fig_spec98S}
\end{wrapfigure}

  In the absence of CSM, the emergence of the shock is prompt and the breakout signal is short-lived (its duration as seen by a distant observer is set by the light-crossing time across the star $R_\star/c$, thus less than an hour in a red-supergiant star). If CSM is present at the progenitor surface, the shock-breakout duration is enhanced, potentially up to days and weeks (Figs~\ref{intsn_fig_obs_div}--\ref{intsn_fig_lbol}). During that phase, the exploding star has the luminosity of a SN, reaching extraordinary values of 10$^8$ to 10$^{10}$\,\lsun, but its photosphere is located in the extended, unshocked CSM enshrouding the star (the photosphere initially migrates outwards with the ionization front caused by the radiative precursor but eventually settles at a fixed radius, which can be many $R_\star$ depending on the CSM extent and density). During that phase, the spectral signatures are thus not broad, testifying for the large velocity of the outer ejecta layers in normal (noninteracting) SNe, but very narrow and reflecting the much lower velocity of the CSM. In practice, the radiation from the shock photoionizes the CSM and increases its optical depth. As photons scatter with free electrons in this CSM (because of their low mass, electrons move with a characteristic velocity $V_{\rm th,e}$ of $5.5 \times 10^5$\,\ms\ at 10,000\,K), their wavelength is modified at each scattering by about $(1 \pm V_{\rm th,e}/c)$ (or in velocity space by $V_{\rm th,e}$). Because the CSM is optically thick, photons scatter multiple times and accumulate such wavelength (or velocity) shifts in a random manner. While the effect is not obvious for continuum photons, it leads to the notorious spectral line profiles of Type IIn SNe characterized by a symmetry relative to the rest wavelength, a narrow line core, and broad wings that extend to a few 10$^6$\,\ms\ (see data and model at 4\,d in Fig.~\ref{intsn_fig_spec98S}).
    
During this phase, the unshocked but ionized CSM is relatively hot (of order 10\,000\,K if the CSM is extended but many 10\,000\,K if the CSM is more compact). Hence, the bulk of the radiation emerges in the ultraviolet range initially and is thus largely missed from ground-based observatories.

\subsection{The `Cold-dense shell' phase}\label{intsn_cds}

With time passing, the SN shock progresses outward through the CSM and suffers some deceleration. Material from the ejecta are slowed down by the CSM and the CSM is accelerated by the ejecta, and thus material from both the ejecta and the CSM piles up into a dense shell in a process similar to a snow-plow. At this time, a distant observer records a hybrid spectrum made of radiation arising from the unshocked CSM as well as from the underlying, faster moving material. Eventually, the entire CSM is snow-plowed and sits in a narrow dense shell, which, because of its high density cools very efficiently (hence the name `cold dense shell' or CDS). This CDS moves at a high velocity of many 10$^6$\,\ms, which is close to that of the outer ejecta located just interior of the CDS.   

Once all the dense CSM has been swept-up by the shock, the spectrum forms entirely in the CDS, which is initially optically thick and thus absorbs any radiation coming from the underlying ejecta. The very sharp density spike associated with that dense shell corresponds to a fast-moving photosphere with a small density scale height (i.e., $dR/d\ln\rho << R$, where $\rho$ is the density at radius $R$). The spectrum thus exhibits a quasi-featureless continuum with only weak blue-shifted absorptions and essentially no line emission. In observed spectra of high quality, such weak features are clearly detected (see data and model at 25--35\,d in Fig.~\ref{intsn_fig_spec98S}).

\begin{wrapfigure}{R}{0.5\textwidth}
  \vspace{-0.9cm}
\begin{center}
\includegraphics[width=0.48\textwidth]{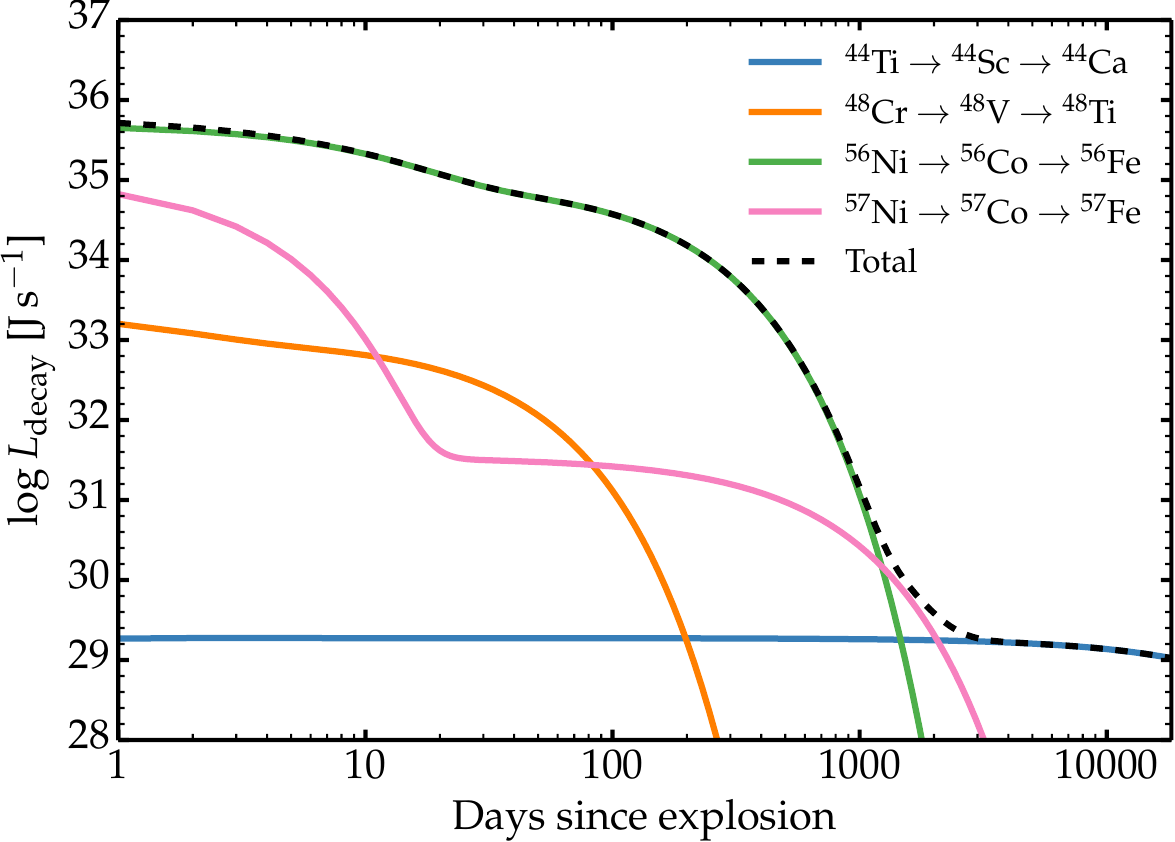} 
\end{center}
\vspace{-0.4cm}
    \caption{Evolution of the radioactive-decay power emitted in a model for a typical core-collapse supernova. We show the four main two-step decay chains that contribute to the total decay power. Because of their different lifetimes, the \nifs-decay chain dominates out to about 2000\,d, followed briefly by that of \nifsev, and eventually by the \tiff\ decay chain out to many years. The initial, undecayed masses for \tiff, \crfe, \nifs, and \nifsev\ are $4.3 \times 10^{-5}$, $1.5 \times 10^{-4}$, $6.3 \times 10^{-2}$, and $2.8 \times 10^{-3}$\,\msun, respectively.}
   \label{intsn_fig_decay}
\end{wrapfigure}

\subsection{The ejecta phase}\label{intsn_ejecta}

From then on, there is little dynamics affecting this complex configuration, and both the CDS and the ejecta evolve ballistically. With expansion, the CDS eventually becomes optically thin and a distant observer can then record photons arising from an extended regions that includes both the ejecta and the CDS. Compared to the previous two phases, the optical spectrum is now more typical of a non-interacting SN (e.g., broad spectral lines are now present), except that some transitions exhibit a much stronger emission with little or no associated absorption. This arises because the emission from the CDS occurs at the highest velocities of all emitting regions (i.e. the CDS sits at the outer edge of the ejecta) and this extra emission compensates for or `fills-in' any absorption that may arise from a spectral feature forming in the underlying ejecta (see data and model at 75--81\,d in Fig.~\ref{intsn_fig_spec98S}).

In reality, the ejecta/CDS configuration is still expanding into the progenitor wind, but now of a much lower density. The CDS may thus be energized by a weak, external shock, which may contribute extra radiation. For example, this extra power may introduce continuum photons that dilute the spectrum and weaken the absorption features. The relative emission from the different regions at this stage are dependent on the external wind density, the CDS mass and velocity, the radiative energy and the \nifs\ in the ejecta.

\subsection{The nebular phase}\label{intsn_late_interaction}

The classification of SNe as interacting rests on the peculiar and rare observation of narrow emission lines in the first spectrum taken of a luminous transient (i.e., luminous enough to warrant a SN name). Because massive stars are notorious for their strong stellar winds,  one would naively expect all massive star explosions to produce interacting SNe. However, only 5\% of core-collapse SNe are classified as Type IIn. The reason is that dense, extended, unshocked CSM must be enshrouding the SN at the time of that first spectral observation in order to flag the transient as a Type IIn SN. With such a strict classification criterion, only the most extreme mass-losing stars can produce SNe IIn. 

\begin{figure}[t]
\centering
\includegraphics[width=0.49\textwidth]{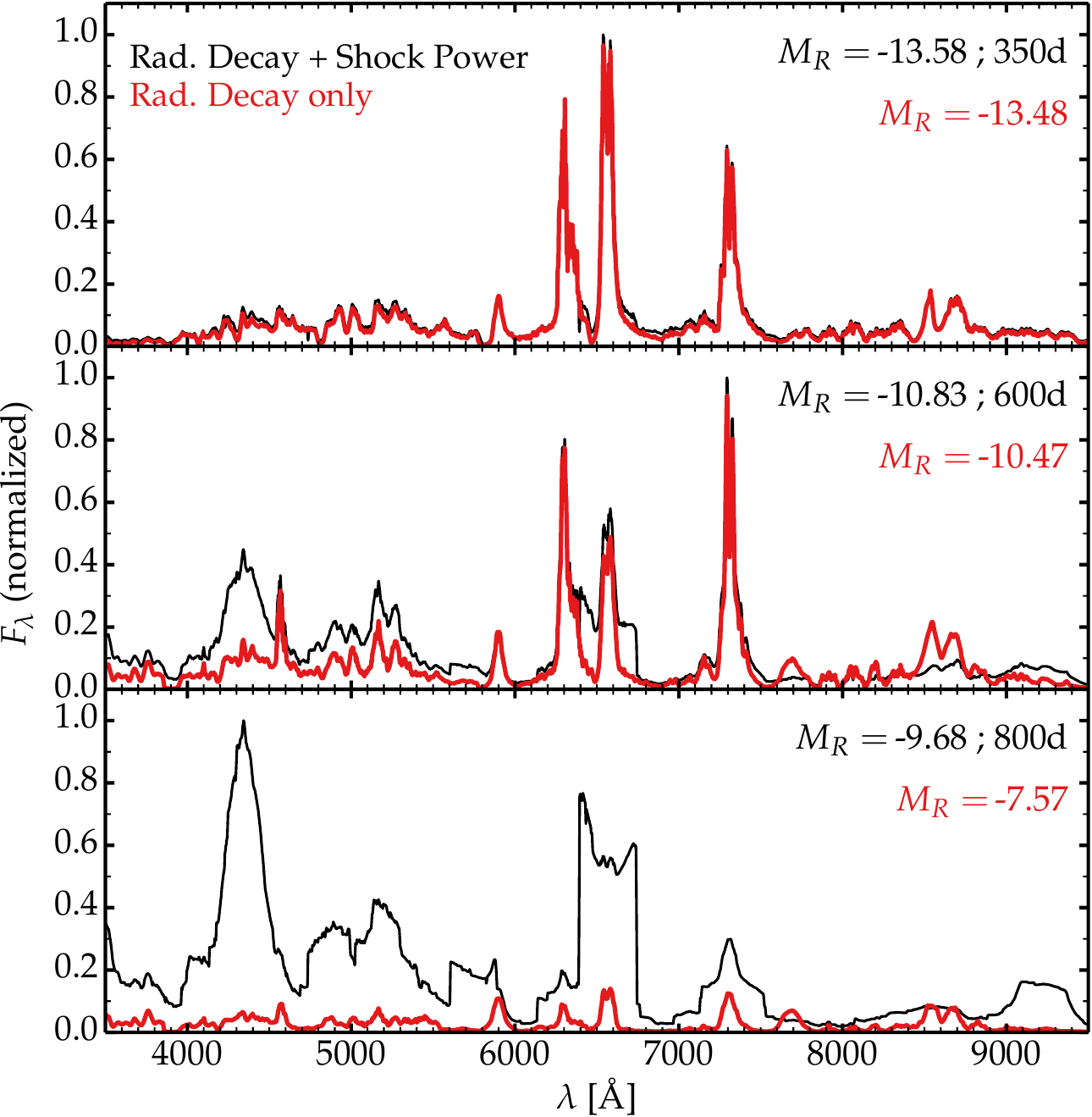} 
\includegraphics[width=0.49\textwidth]{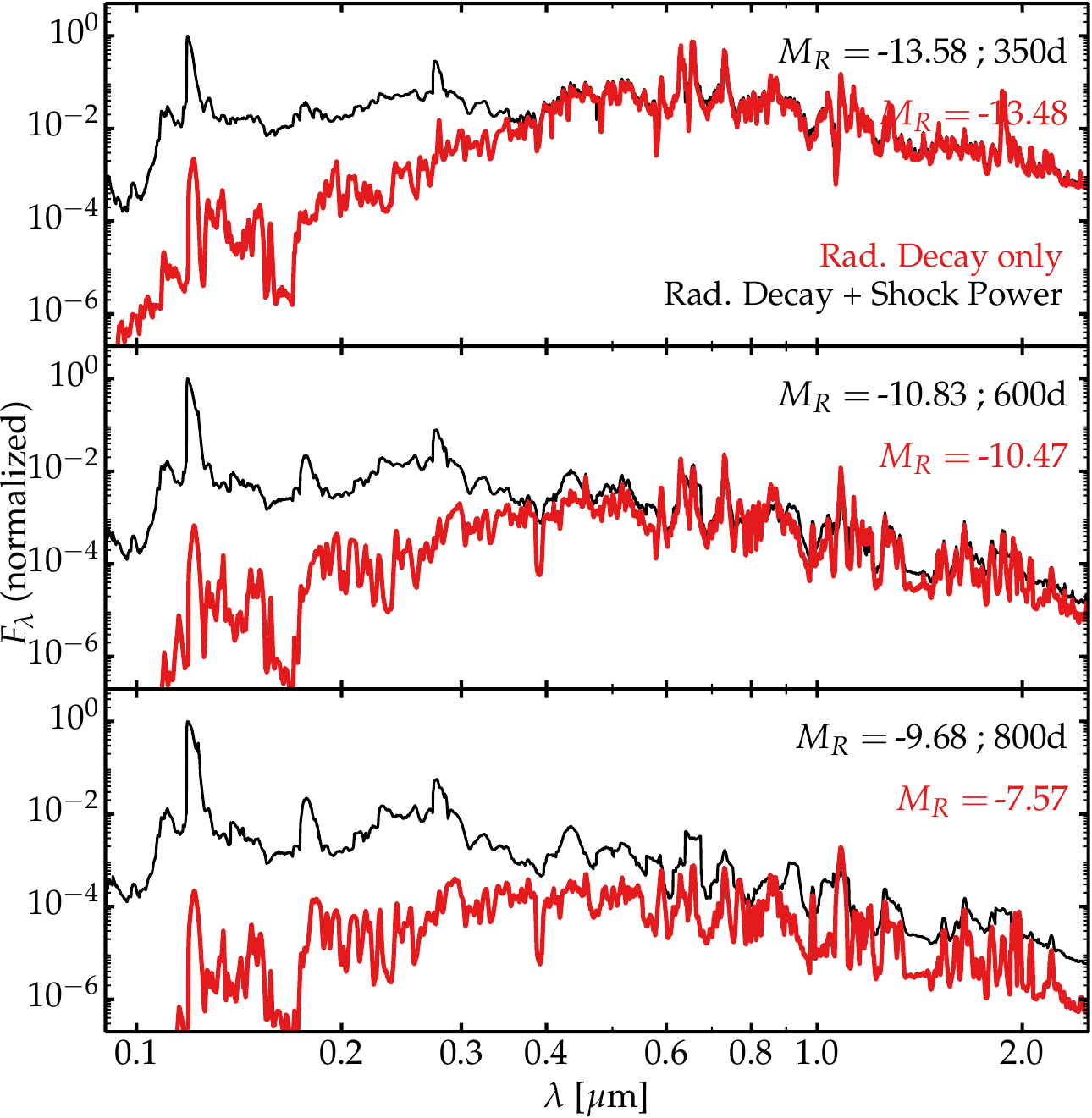} 
    \caption{Illustration of the impact of shock power on the optical (left panel) and ultraviolet-to-infrared (right panel) ranges in an otherwise standard Type II-P SN. While the model powered by radioactive decay shows relatively narrow emission lines that weaken in time, the model counterpart that includes shock power as well stay luminous as time progresses, in particular in the ultraviolet, and exhibits broad, boxy emission lines that form in the outer ejecta.}
\label{intsn_fig_late_spec}
\end{figure}

The difficulty of identifying the signatures of interaction arises in part from the fact that interaction power is only a fraction of the total power radiated by the SN. Indeed, during the photospheric-phase, the ejecta release on a diffusion timescale the energy left over by the shock during its journey through the envelope of the star (this power is sustained only in explosions of extended stars like RSGs) as well as the energy from radioactive decay.  In a standard Type II-P SN, this escaping radiative power during the first few months is about 10$^{35}$\,\js, but as the ejecta expand and turn nebular, this power drops by a factor 100 or more and arises typically only from radioactive decay. Because this power source declines exponentially (Fig.~\ref{intsn_fig_decay}), the conditions for detecting signatures of ejecta interaction with CSM become increasingly favorable as time passes. Put more broadly, ejecta interaction with CSM represents a long-lived power source that may make SNe shine for decades after explosion as they slowly turn into a SN remnant \citep{black_13by_17}.

Numerical simulations can help in predicting the radiative signatures of ejecta influenced by these various energy sources and confrontation to observations can then be used to constrain the magnitude of each source. Fig.~\ref{intsn_fig_late_spec} illustrates the simulated evolution of a Type II SN ejecta from 350 to 800\,d after explosion in both the optical (left panel) and across the ultraviolet, optical and near-infrared (right panel; log-log plot). This specific example uses as reference the ejecta of a SN II-P model powered at such late times by the decay of \cofs\ (red curve; Fig.~\ref{intsn_fig_decay}), whereas in the model counterpart (black curve), an additional constant shock power of 10$^{33}$\,\js\ is injected in the outer ejecta near 1.1$\times$10$^7$\,\ms\ \citep{dessart_late_23}. This extra power switches from subdominant to dominant at $\sim$\,600\,d. In the optical, this evolution is marked by the progressive strengthening of broad, boxy profiles in the optical (e.g., H$\alpha$), powered by the ejecta/CSM interaction, and the weakening of the decay-powered inner, metal-rich ejecta (where forbidden lines like \oidoub\ form). Furthermore, in the model with interaction power, the ultraviolet contains the bulk of the escaping flux, which becomes progressively channeled into a few strong emission lines like Ly$\alpha$. 
Although rarely obtained, panchromatic observations covering from the ultraviolet to the optical confirm this property (e.g., for Type IIn SNe 1998S, 1995N, or 2010jl, but also in the Type IIb SN\,1993J; \citealt{fransson_cno_93J_98S_05,fransson_95N_02,fransson_10jl}).

\section{Understanding the diversity of interacting, Type IIn SNe}
\label{intsn_iin_diversity}

The observed photometric and spectroscopic evolution of  interacting SNe, for which we show a small but representative sample in Fig.~\ref{intsn_fig_obs_div}, implies a diversity in the properties of the CSM (and of the ejecta, though probably to a lesser extent) -- see also Section~\ref{intsn_rhd}. We now study the signatures of the three main categories of interacting SNe, which we associate with short-lived, long-lived, and quenched interactions.

  \subsection{Short-lived interactions: SN\,2013fs analogs} 

\begin{figure*}
\vspace{-0.1cm}
\begin{center}
\includegraphics[width=0.48\textwidth]{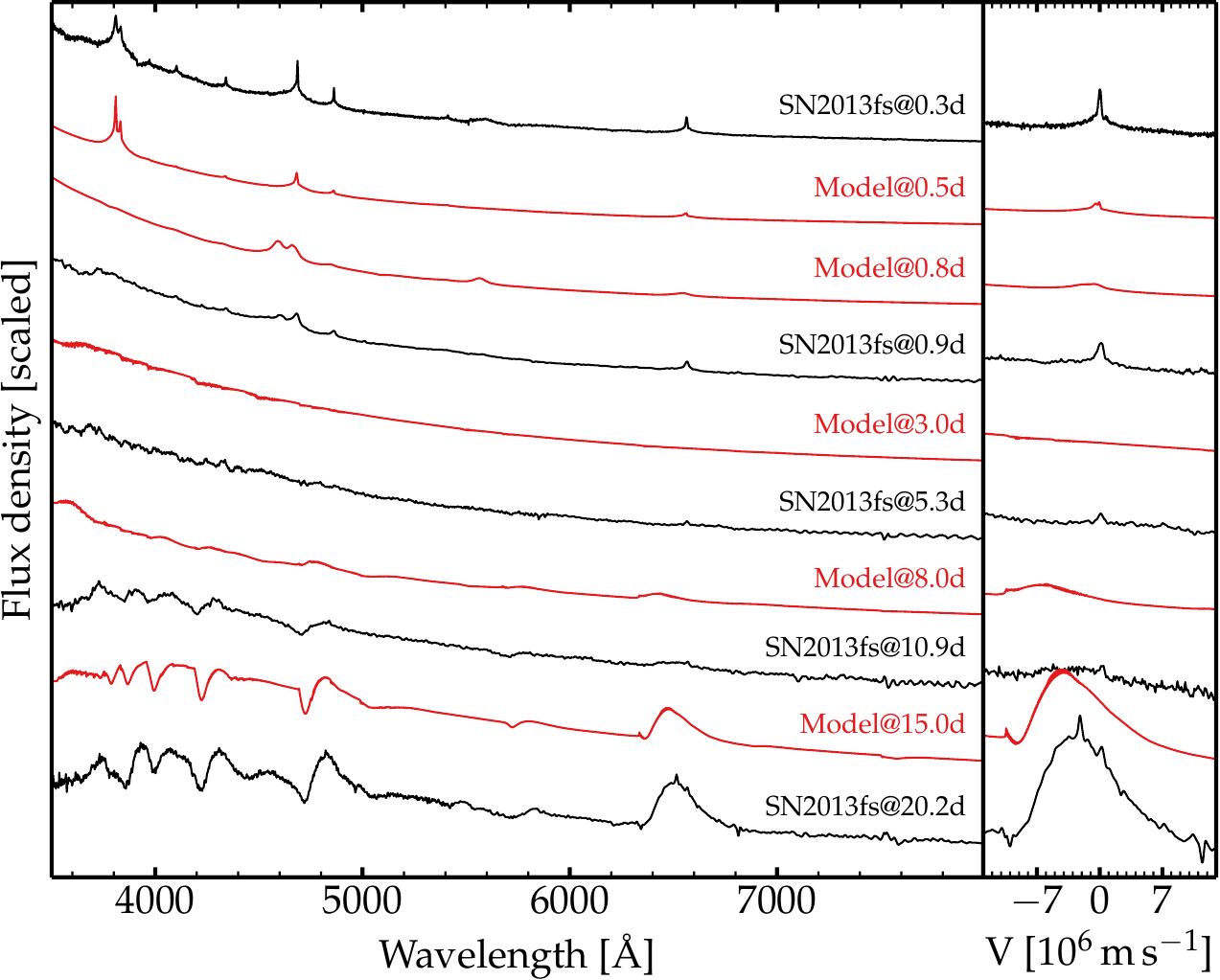} 
\includegraphics[width=0.48\textwidth]{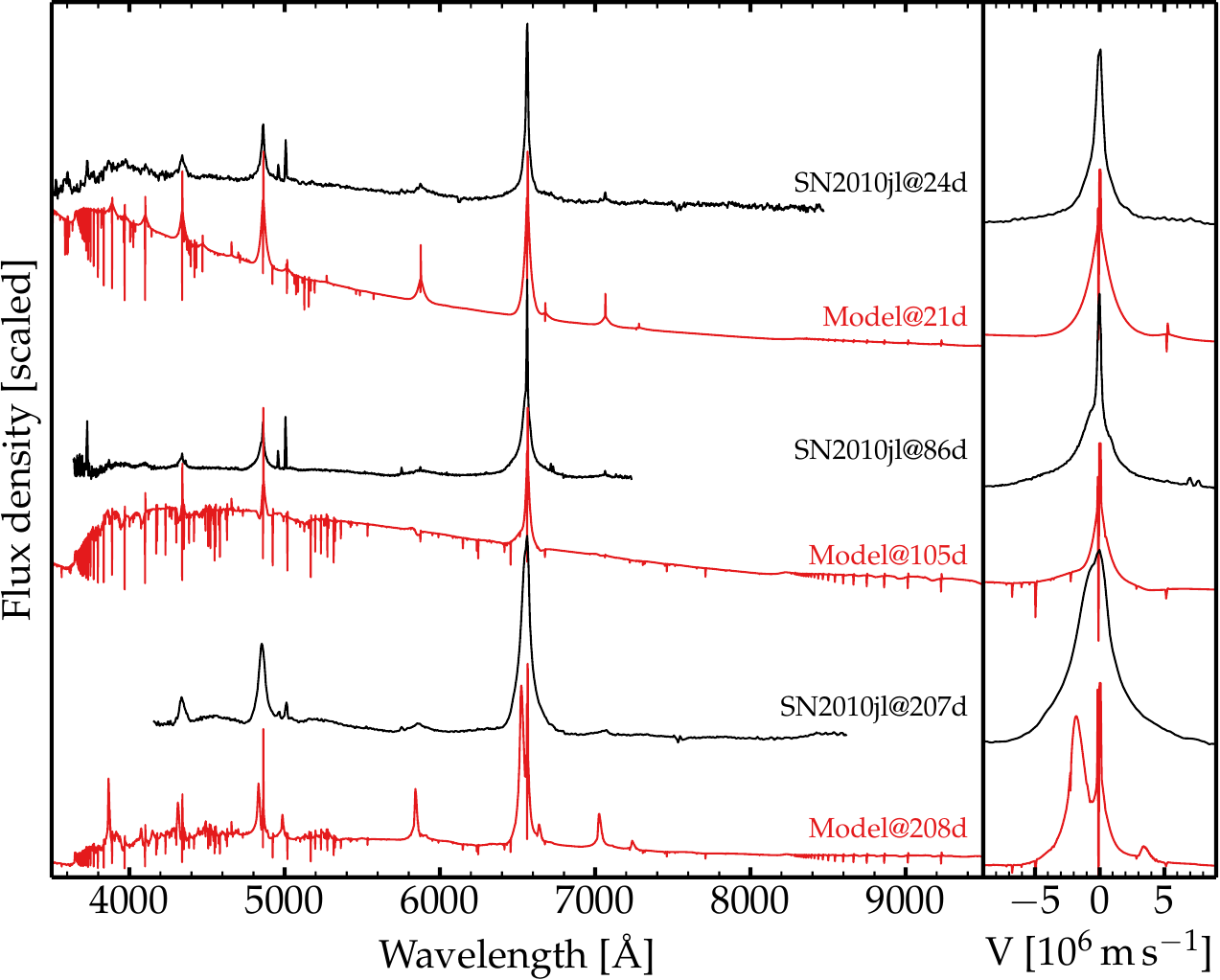} 
\end{center}
\vspace{-0.4cm}
    \caption{Same as Fig.~\ref{intsn_fig_spec98S}, but now showing the spectral evolution of the protypical short-lived interaction transient SN\,2013fs (left; model results are from \citealt{d17_13fs}) and long-lived interaction transient SN\,2010jl (right; model results are from \citealt{D15_2n}).}
   \label{intsn_fig_spec13fs_10jl}
\end{figure*}

By short-lived interactions, we mean events in which the signatures of interaction are only temporary, with IIn-like spectral signatures present at discovery but fading and disappearing in a matter of hours or days, after which the SN slowly evolves into a standard-looking Type II SN. Physically, this corresponds to a case where the CSM mass is much smaller than the ejecta mass (say 1\% of $M_{\rm ej}$) and where the dense CSM extends out to 1--10 times the progenitor surface radius. Quantitatively, the IIn-like signatures survive for a duration $(R_{\rm CSM}  - R_\star) / V_{\rm sh}$, where $R_{\rm CSM}$ corresponds to the outer edge of the dense CSM and $V_{\rm sh}$ is typically about 10$^7$\,\ms\ (complicated processes influence the exact value of  $V_{\rm sh}$, which is affected by radiation leakage and interaction with CSM). SN\,1998S, which was discussed in Section~\ref{intsn_ref_case} is a clear member of this class, with IIn signatures (i.e., associated with H\one, He\two, and N\three\ lines) that last for 1-2 weeks before the SN enters the `CDS phase'. When IIn signatures are present for only a day after shock breakout, the CSM is less massive, less extended (of order 10$^{12}$\,m), the luminosity boost is smaller (less kinetic energy is extracted by the CSM) and the spectral signatures are of much higher ionization (the CSM traps a large energy in a smaller volume at smaller radii and is therefore hotter and more ionized). SN\,2013fs is the prototype of that case \citep{yaron_13fs_17}, which may be very frequent but was observed only once because it requires a discovery within hours of shock breakout and immediate, hour-cadence spectral monitoring (see left panel of Fig.~\ref{intsn_fig_spec13fs_10jl}). 

The origin of this CSM is unclear today. Possible scenarios invoke an enhanced mass loss related to interior instabilities taking place in the ultimate phases of massive star evolution (see Section~\ref{intsn_wave}), a universal mass overloading of the direct stellar environment related to the combination of inefficient wind driving and surface instabilities (e.g., pulsations, convection; Section~\ref{intsn_wind}), or mass loss phenomena inherent to case C binary systems (Section~\ref{intsn_bin}). The frequency of SNe IIP with early-time IIn signatures is, however, relatively high \citep{strotjohann_presn_21,bruch_csm_23} and therefore requires a phenomenon that is both robust and universal.
  
  \subsection{Long-lived interactions: SN\,2010jl analogs}

By long-lived interactions, we mean events in which the signatures of interaction are essentially present at all times. The exploding star plows through CSM as the SN shock continuously encounters `fresh' material to interact with, and the event never looks like a standard SN. These events essentially remain in the `CSM-phase' at all epochs. A prototype of this class of long-lived interactions is the super-luminous Type IIn SN\,2010jl, which was characterized by a high brightness phase (Fig.~\ref{intsn_fig_obs_div}) and IIn signatures that lasted for an entire year. The CSM is in this case very extended, covering out to 10$^{14}$\,m and may represent about 10\,\% of the ejecta mass, allowing considerable kinetic energy to be extracted and radiated away. Such events suggest a prolonged phase of intense mass loss in the decades leading to core collapse. 

Historically, such phenomena have been associated with the eruptive mass loss observed in high-mass massive stars like $\eta$ Carina (with as much as 10-20\,\msun\ of material lost in the eruption; \citealt{smith_etacar_03}), although the CSM mass in SN\,2010jl is estimated to be only a few solar masses \citep{fransson_10jl,D15_2n}. Further evidence is given by the spectral evolution which indicates a progressive broadening of emission lines during the first year  (see right panel of Fig.~\ref{intsn_fig_spec13fs_10jl}). This suggests that the ejecta eventually emerged from the dense CSM region, and that at that time it had retained a significant kinetic energy. The extended optically-thick CSM represents a huge volume which tends to yield a smaller temperature and ionization than in the more compact configurations associated with short-lived interactions.

\begin{wrapfigure}{R}{0.5\textwidth}
\vspace{-0.9cm}
\begin{center}
\includegraphics[width=0.48\textwidth]{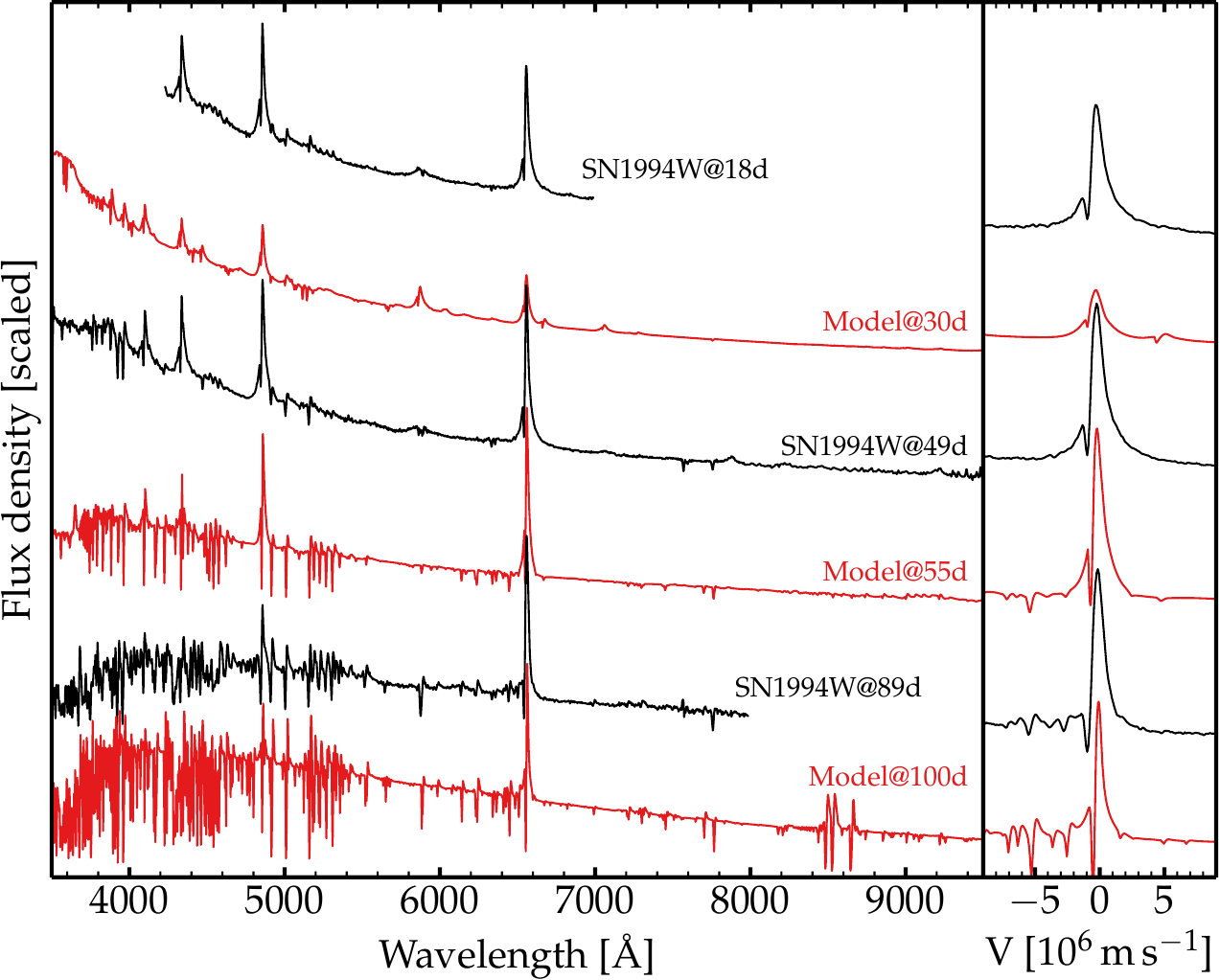} 
\end{center}
\vspace{-0.4cm}
    \caption{Same as Fig.~\ref{intsn_fig_spec98S}, but now showing the spectral evolution of the quenched interaction SN\,1994W, which exhibits narrow spectral lines at all times.}
   \label{intsn_fig_spec94W}
\end{wrapfigure}

  \subsection{Quenched interactions: SN\,1994W analogs}

Historically, interacting SNe were viewed as a class of events in which an energetic, massive ejecta encountered a moderately dense wind, in other words a very unequal confrontation with $M_{\rm ej} >> M_{\rm CSM}$ and in which the ejecta `win' by sweeping up the CSM and remain largely unaffected during the interaction. In quenched interactions, the relation is reversed and we have $M_{\rm ej} << M_{\rm CSM}$ so that the CSM has the ability to decelerate considerably the ejecta (i.e. extracting most of its kinetic energy) and to radiate away that energy to produce a luminous transient. The conversion efficiency of ejecta kinetic energy to radiation energy is no longer a few percent as in short-lived interactions like SN\,2013fs, but can instead reach up to 100\% (see also Table~\ref{tab_rhd} and Fig.~\ref{intsn_fig_lbol}).

A prototype for these quenched interactions is SN\,1994W \citep{sollerman_etal_94w,chugai_94w_04} which exhibited a 100\,d-long high-brightness phase with a luminosity of order 10$^9$\,\lsun, followed by a precipitous drop to a low brightness, and relatively cool spectra with lines that remained narrow at all times (Fig.~\ref{intsn_fig_spec94W}). Narrow lines at early times reflect a spectrum formation in the slow unshocked CSM but in interacting SNe, this phase is generally followed by the emergence of broad lines from the underlying ejecta. The absence of such broad lines at late times indicates that the ejecta have been totally decelerated, as obtained in simulations that invoke a configuration with $M_{\rm ej} << M_{\rm CSM}$ \citep{D16_2n}.
This type of configuration is likely at the origin of the most luminous interacting SNe like 2006gy \citep{smith_06gy_08,jerkstrand_06gy_20} since their time- and wavelength-integrated luminosity reaches the typical total energy available in stellar explosions (i.e., 10$^{44}$\,J). 

\section{Other peculiar interacting supernovae}
\label{intsn_other}

The most frequent interacting SNe are of Type IIn and belong to the various classes discussed in Section~\ref{intsn_iin_diversity}, with prototypical events SNe 1998S/2013fs, 2010jl, and 1994W. The diversity of interacting SNe is, however, much greater and includes more exotic albeit rarer events. 
 
 \subsection{Type Ibn SNe}
 
\begin{wrapfigure}{R}{0.5\textwidth}
\vspace{-1.cm}
\begin{center}
\includegraphics[width=0.48\textwidth]{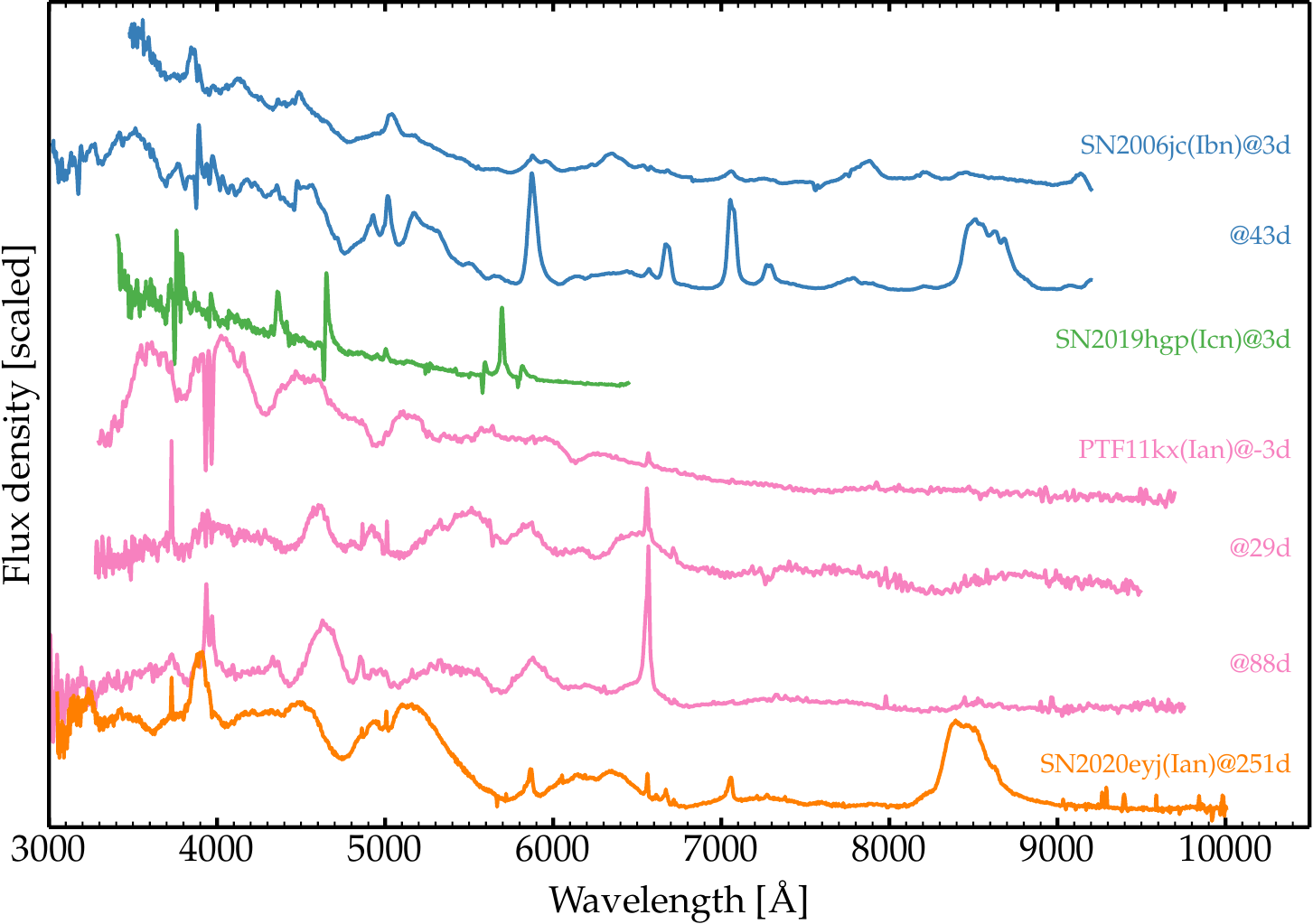} 
\end{center}
\vspace{-0.4cm}
    \caption{Illustration of spectral properties for rare and peculiar interacting SNe with Types Ian, Ibn, and Icn (see Section~\ref{intsn_other} for discussion).}
   \label{intsn_fig_noh}
\end{wrapfigure}

 Type Ibn SNe, which were briefly mentioned in the introduction, are luminous transients whose spectra exhibit He\one\ lines that are persistent and relatively narrow (width of order 10$^6$\,\ms; Fig.~\ref{intsn_fig_noh}). The first object of this type was SN\,2006jc \citep{foley_06jc_07,pasto_06jc_07}, but the sample now includes several dozen objects \citep{hosseinzadeh_ibn_17}. Their photometric evolution is rapid, with a fast rise to an $R$-band maximum of about $-$19\,mag in about a week and a fast decline thereafter, typically by three magnitudes per month, although there is much scatter. This fast-rise and fast-decline are caused by the small mass involved in those interactions relative to SNe IIn (there is about one solar mass of He-rich material in a massive star). The electron-scattering optical depth is also reduced by the larger abundance of metals (i.e., the lack of H) and the relatively high ionization potential of He\one. The most spectacular feature of SNe Ibn is their unique spectral characteristics dominated after maximum by a strong emission from a forest of Fe\two\ lines and few isolated lines of He\one, C\two, and Mg\two. Although SNe Ibn are complex phenomena to interpret, numerical simulations suggest that 3--4\,\msun\ He stars formed in interacting binaries may be at the origin of these events (see, e.g., \citealt{dessart_ibn_22}; \citealt{wu_fuller_22}).
 
\subsection{Type Icn SNe}
  
  A new and hence rare type of transients was recently discovered that exhibit a similar light curve to Type Ibn SNe but with spectra dominated by lines of C and O and none of H nor He (e.g., SNe 2019hgp and 2021csp; \citealt{galyam_19hgp_22}, \citealt{perley_21csp_22}). The discovery spectra exhibit Doppler-broadened lines indicating velocities of order 10$^6$\,\ms, reminiscent of what is observed in Wolf-Rayet stars. The origin for this dense and extended material is currently a mystery, and it is not clear how these SNe Icn stand relative to SNe Ibn, Ib, and Ic in terms of progenitor and explosion properties.
 
\subsection{Type Ian SNe}
 
 So far in this chapter, we have considered the interaction between ejecta and CSM in the context of massive stars. Signatures of interaction may, however, arise also in Type Ia SNe, and thus likely in association with the explosion of a white dwarf star. Evidence comes from transients similar in appearance to standard Type Ia SNe apart from the presence of narrow and persistent spectral lines of H\one\  (e.g., SNe\,2002ic, \citet{hamuy_02ic_03}; PTF11kx,  \citet{dilday_11kx_12}; see also \citealt{silverman_ian_13}) or of He\one\ (SN\,2020eyj; \citealt{kool_iacsm_23}). Finding CSM around SNe Ia is not strictly surprising since their white dwarf progenitors, residing in a binary system, lose several solar masses during their lifetime, but they do so over billions of years of evolution and thus such CSM should be very distant (i.e., parsec scale rather than 10$^{13}$--10$^{14}$\,m) at the time of explosion. The scarcity of SNe Ian suggest that these special configurations are indeed very rarely encountered in exploding white dwarfs.
 
\section{Dust formation}
  \label{intsn_dust}
  
 The slowly-expanding, metal-rich inner ejecta of core-collapse SN ejecta are the site of dust formation on year timescales, which may be diagnosed from an anomalously large infrared luminosity (identifying an anomaly is not trivial since all transients may form dust) or from the asymmetry of line profiles (i.e., dust attenuation is greater for escaping photons that originated from the back side of the ejecta). Such dust has been identified in the inner ejecta of SN\,1987A (e.g., with observations by the satellite Herschel; \citealt{matsuura_87A_11}) and through profile asymmetry \citep{lucy_dust_89}. In addition, at late times, the potential interaction of the ejecta with CSM may also lead to the formation of dust. This seems to occur in bona-fide interacting SNe like 2010jl \citep{fransson_10jl}, but also across a wide range of historical Type II-P or II-L SNe \citep{niculescu_duvaz_dust_22,shahbandeh_jwst_23}. This topic is fast evolving because of the abundance infrared observations performed by the James Webb Space Telescope.
 
\section{Concluding remarks} 
\label{intsn_conc}

While the dynamical and radiative phenomena associated with the interaction between ejecta and CSM are relatively well understood, the community faces a considerable challenge in elaborating a comprehensive, global understanding of interacting SNe. For the origin of the CSM, a set of physical processes have been identified in single and binary massive stars across a wide range of masses but connecting any of these to a specific given event as well as explaining their relative rates is difficult. It is clear that interacting SNe are too numerous to be explained exclusively by eruptions and explosions from the most massive stars, and the majority of interacting SNe may in fact arise from  lower mass massive stars, in particular those residing in binary systems. Much work remains to be done.

In spite of these difficulties and uncertainties, the community has undoubtedly made considerable progress in the last decade, for example in documenting the occurence of interactions not only early at the time of shock breakout but also late when the ejecta slowly transition to a SN remnant. Observations in the X-rays, ultraviolet, infrared, and radio wavelengths have complemented those in the optical, and new types of interactions have been identified. The next decade will bring refined information on interacting SNe with the deep optical survey conducted with the Vera Rubin Observatory \citep{LSST}, the high-cadence ultraviolet survey with ULTRASAT \citep{ULTRASAT}, and the ultraviolet photometric and spectroscopic monitoring with UVEX \citep{uvex}. In parallel, improved realism of radiation-hydrodynamics and radiative transfer simulations will assist with the interpretation of this incoming wave of observations.

\bibliographystyle{Harvard}
\bibliography{interacting_supernovae}

\end{document}